\documentclass[12pt,aps,preprint,floatfix,pra]{revtex4}
\usepackage[utf8]{inputenc}
\usepackage{amsmath}
\usepackage{tikz}
\usepackage{tkz-euclide}
\usepackage{dsfont}
\usepackage{amssymb}
\usepackage{amsmath}
\usepackage{mathrsfs}
\usepackage{appendix}
\usepackage{bbold}
\usepackage{color}
\usepackage{xcolor} 
\usepackage{braket}
\usepackage[hyphenbreaks]{breakurl}  
\usepackage{microtype}  
\usepackage{hyperref}
\usepackage[english]{babel}
\usepackage{breakurl}
\usepackage[utf8]{inputenc}
\usepackage{comment}
\usepackage{textcomp}          
\usepackage{amsmath}
\usepackage{bm}
\DeclareMathOperator{\Var}{Var} 

\DeclareMathOperator{\Cov}{Cov}
\DeclareMathOperator{\Tr}{Tr}

\renewcommand{\Re}{\operatorname{Re}}
\renewcommand{\Im}{\operatorname{Im}}
\def\ie{\begin{equation}\begin{aligned}}
\def\fe{\end{aligned}\end{equation}}

\begin{document}

    \title{Time uncertainty and fundamental sensitivity limits in quantum sensing: application to optomechanical gravimetry}
	
	\author{Salman Sajad Wani}
	\affiliation{Qatar Center for Quantum Computing, Collage of Science and Engineering, Hamad Bin Khalifa University, Doha, Qatar} 
	\author{Saif Al-Kuwari}
	\affiliation{Qatar Center for Quantum Computing, Collage of Science and Engineering, Hamad Bin Khalifa University, Doha, Qatar}
	\author{Arshid Shabir}
	\affiliation{Canadian Quantum Research Center, 460 Doyle Ave 106, Kelowna, BC V1Y 0C2, Canada}
	\author{Paolo Vezio}
	\affiliation{Dipartimento di Fisica e Astronomia, Universit\'a di Firenze, Via Sansone 1, I-50019 Sesto Fiorentino (FI), Italy.}
	\author{Francesco Marino}
	\affiliation{CNR-Istituto Nazionale di Ottica and INFN, Via Sansone 1, I-50019 Sesto Fiorentino (FI), Italy.}
	\author{Mir Faizal}
	\affiliation{Canadian Quantum Research Center, 460 Doyle Ave 106, Kelowna, BC V1Y 0C2, Canada}
	\affiliation{Irving K. Barber School of Arts and Sciences, University of British Columbia Okanagan, Kelowna, BC V1V 1V7, Canada.}
	\affiliation{Department of Mathematical Sciences, Durham University, Upper Mountjoy, Stockton Road, Durham DH1 3LE, UK}
	\affiliation{Computational Mathematics Group, Faculty of Sciences, Hasselt University, 
Agoralaan, Gebouw D, 3590 Diepenbeek, Belgium}

	\begin{abstract}	
High-sensitivity accelerometers and gravimeters, achieving the ultimate limits of measurement sensitivity are key tools for advancing both fundamental and applied physics. While numerous platforms have been proposed to achieve this goal, from atom interferometers to optomechanical systems, all of these studies neglect the effects of intrinsic quantum uncertainty in time estimation. 
Starting from the Hamiltonian of a generic linear quantum sensor, we derive the two-parameter quantum Fisher information matrix and establish the corresponding Cram\'er-Rao bound, treating time as an uncertain (nuisance) parameter. Our analysis reveals a fundamental coupling between time and signal estimation that inherently degrades measurement sensitivity, with the standard single-parameter quantum limit recovered only at specific interrogation times or under special decoupling conditions. We then apply these results to an optomechanical gravimeter and explicitly derive an optimal decoupling condition under which the effects of time uncertainty are averaged out in a continuous measurement scheme. Our approach is general and can be readily extended to a broad class of quantum sensors. 
\end{abstract}
\maketitle

\section{Introduction}\label{section_i}
\text

Quantum-limited sensing, aiming for the best possible precision allowed by quantum mechanics, and quantum-enhanced sensors, which utilizes quantum resources top improve precision beyond classical limits, find applications in numerous areas of fundamental and applied research. In particular, precise measurements of the gravitational acceleration $g$ is crucial for numerous applications in geoscience and space exploration. For instance, local gravity variations due to mass redistribution driven by climate change have been mapped with the GRACE satellite \cite{GRACE1,GRACE2} and the Juno spacecraft mission \cite{JunoGravity} characterized gravitational field of Jupiter, providing important information on its formation and interior structure. At a fundamental level, high-precision gravimetry offers the potential to test general relativity \cite{ciufolini2016,kasevich}, advance relativistic quantum metrology \cite{fuentes14,fuentes19,fuentes23}, and detect possible deviations from Newtonian gravity \cite{mazumdar12}.
	
In this context, quantum gravimeters, devices that exploit quantum phenomena to enhance measurement sensitivity, are at the forefront of gravity measurement technology. Systems prepared in entangled or squeezed quantum states can achieve measurement uncertainties that scale as $1/N$ (Heisenberg limit), rather than the classical \(1/\sqrt{N}\) scaling of the standard quantum limit, where $N$ is the number of probes. This quantum advantage has been proposed for gravimetric applications through the incorporation of quantum state engineering techniques into sensor design \cite{Montenegro2025}. However, as emphasized in Ref.~\cite{zwierz}, the Heisenberg limit should be understood more generally, not merely as a $1/N$ scaling law, but as the ultimate precision bound determined by the total available physical resources, which can be expressed through the variance of the generator of the parameter-encoding transformation, consistent with the quantum speed limit governing the system’s evolution.

These ideas have inspired a number of proposals and experimental efforts aimed at developing sensors able to operate at the ultimate sensitivity limits imposed by quantum mechanics. A notable example is atom interferometry, in which ultracold atoms in free fall form matter-wave interferometers \cite{Rosi2014}. In these devices, gravitational acceleration induces a phase shift in the interference pattern, enabling measurements of $g$ with remarkable precision \cite{Asenbaum2020}. Early demonstrations using atom interferometers have shown that their performance is competitive with or even superior to that of conventional gravimetric systems \cite{tinoreview}. For example, sensitivities as low as \(\Delta g = 4.3 \times 10^{-9}\)~ms\(^{-2}\) has been achieved using cold atoms in a two-dimensional magneto-optical trap \cite{AtomInterferometry1}, while even greater sensitivity of \(\Delta g = 7.8 \times 10^{-10}\)~ms\(^{-2}\) has been reached in a similar investigation using Bose-Einstein condensates \cite{BECInterferometry}. Moreover, a dual-species atom interferometer allowed to measure the relative acceleration between $^{85}$Rb and $^{87}$Rb atoms at the level of $10^{-12}$g \cite{kasevich}. 
	
More recently, proposals based on magnetically levitated spheres \cite{Lewandowski2021, Monteiro2017} have suggested the possibility of achieving sensitivities around \(2.2 \times 10^{-9}\)~ms\(^{-2}\)~Hz\(^{-1/2}\) \cite{Magnetomechanics,Magnetomechanics1}. Optically levitated nanoparticles, which additionally exploit optomechanical coupling within high-finesse optical cavities, can potentially reach even better sensitivities on the order of \(\Delta g \sim 10^{-15}\) m/s\(^2\) \cite{NonlinearGravimetry2018}. In these systems, the levitated nanosphere works as an almost ideal test mass. In the absence of a mechanical clamping mechanism, their motion can be highly isolated from environmental decoherence, potentially resulting in long coherence times for matter-wave interferometry and enabling free-fall experiments. Consequently, these systems offer a promising platform for high-sensitivity measurements of gravitational force.
	
The best fundamental sensitivity $\Delta g$, defined as the standard deviation of a gravimetric measurement, can be directly inferred from the system dynamics by evaluating the Fisher information, which provides a natural lower bound on the variance of an unknown parameter (in this case, $g$) via the Cram\'er Rao inequality \cite{Montenegro2025, NonlinearGravimetry2018}. In these works, as well as in previous studies (see e.g. \cite{Rosi2014, Asenbaum2020, tinoreview}), the sensitivity is evaluated using the single-parameter Fisher information with respect to $g$, thereby implicitly assuming time as a classical variable. 
On the other hand, quantum speed limits reveal a fundamental connection between energy and time uncertainty, establishing a fundamental bound on the minimum time required for a quantum system to evolve between two distinguishable states. The Mandelstam-Tamm bound quantifies this minimal time, showing that it is inversely proportional to the energy uncertainty \cite{MandelstamTamm1945}. Helstrom later showed that the precision with which time can be estimated is fundamentally limited by the energy variance, via the quantum Cram\'er-Rao bound (QCRB) \cite{Helstrom1976}, which relates the estimation error to the inverse of the quantum Fisher information (QFI). From a geometric perspective, Braunstein and Caves showed that this bound can be understood in terms of the Bures (statistical) distance between quantum states \cite{Braunstein1994}.

In quantum parameter estimation theory, time must be formally treated as an indirect parameter that can only be probed through the system's Hamiltonian using the QFI. There is an extensive literature on the metrology of time, mainly focusing on estimating time or proper‐time intervals using quantum systems \cite{ahmadi,lind}, or on improving clock stability via entanglement \cite{colombo}. While our analysis shares conceptual connections with these works—such as the role of the QFI for time estimation and its relation to quantum speed limits—our focus is fundamentally different. Here, time is not the quantity of interest, but a parameter that enters the dynamics and unavoidably fluctuates due to quantum uncertainty, thereby affecting the precision with which a signal (e.g., acceleration) can be estimated. Our aim is therefore to analyze how the coupling between the generator of signal encoding and the generator of time evolution impacts the achievable sensitivity bounds.

{This approach is conceptually distinct from jointly estimating the evolution time and the signal parameter, i.e. from the simultaneous reconstruction of both quantities, and reflects realistic sensing scenarios in which the interrogation time is not directly observed, but enters indirectly through the system Hamiltonian. Treating time as a nuisance parameter is therefore appropriate for modeling measurement settings in which time influences the probe dynamics, but it is not the target of estimation.}
{The strategy is to formulate the problem as a two-parameter quantum estimation task, in which both the signal parameter $a$ and the evolution time $\tau$ enter the Hamiltonian and jointly determine the quantum state, and then to eliminate $\tau$ by optimizing over it. As shown below, this is accomplished by constructing the full two-parameter QFI matrix and extracting the effective Fisher information for $a$ via the Schur complement of the nuisance block. }

On this basis, we extend previous analyses on the ultimate precision limits in gravimetry by treating both time and gravitational acceleration as intrinsic parameters within the system's quantum Hamiltonian. Starting from the Hamiltonian of a generic linear quantum sensor, with time treated as a nuisance parameter, we derive the two-parameter QFI matrix and establish the corresponding QCRB for parameter estimation. Our results reveal that time uncertainty inherently couples to the parameter of interest, thereby degrading the achievable sensitivity, except at specific interrogation times. We then apply our findings to a cavity-optomechanical system, which has recently been proposed as a promising platform for high-sensitivity measurements of gravitational acceleration \cite{Armata,NonlinearGravimetry2018,qvarfort2021optimal}. We explicitly derive the ultimate sensitivity limits and identify an optimal decoupling condition under which the estimation of gravitational acceleration is, on average, decoupled from time uncertainty, thereby restoring the usual (single-parameter) Heisenberg-limited sensitivity in a continuous measurement scheme. Finally, we calculate the classical Fisher information (CFI) for homodyne detection and show that appropriate measurement strategies could still approach the quantum limit if carefully optimized. While in this work we focus on a specific system, our results are generally applicable to a broad class of quantum-limited sensing platforms, including gravimetry experiments based on atomic interferometers.
	
The paper is organized as follows. In Sec. II we introduce a general model for a quantum sensor governed by a parameter-dependent Hamiltonian and derive the corresponding QFI, establishing the framework for parameter estimation in subsequent sections. In Sec. III, we introduce a prototype Hamiltonian model describing a quantum optomechanical system which, ideally, could reach Heisenberg-limited sensitivity in gravitational acceleration measurements \cite{NonlinearGravimetry2018,qvarfort2021optimal}. In Sec. IV, we derive the full two-parameter QFI matrix for simultaneous estimation of gravitational acceleration and evolution time, and we obtain explicit expressions for the ultimate precision bounds attainable in such system. In Section V, we discuss the CFI matrix associated with homodyne detection. The conclusions are presented in Sec. VI.
	
\section{General formalism}\label{section_ii}
	
We start by considering the Hamiltonian of a generic linear quantum sensor
\begin{equation}
\hat H(a)=\hat H_{0}+ \lambda a \,\hat Q\,,
\label{eq:generic}
\end{equation}
where $\hat{H_{0}}$ governs the dynamics of the sensing element in the absence of external forces. The linear coupling term $\lambda a\,\hat{Q}$ describes any interaction (at first order) in which a parameter of interest $a$ is encoded into an observable $\hat{Q}$. The physical meaning of $a$ and $\hat{Q}$ depends on the specific system: in the specific case of our optomechanical gravimeter, $a=g$ is the acceleration due to gravity, $\hat{Q}$ is the position operator $\hat{x}$ and $\lambda$ is the inertial mass \cite{NonlinearGravimetry2018}. Most quantum accelerometer/gravimeter e.g., based on trapped atoms, superconducting circuits, or photonic-crystal devices, can be described in terms of Eq.~\eqref{eq:generic}. More generally, for a charged particle $q$ in an electric field $E$ the interaction term takes the form $\lambda a \hat{Q}= q E \hat{x}$, as in trapped-ion or nano-electromechanical electrometers, where the field induces a displacement of the charged oscillator \cite{gilmore,bonus}. Similarly, for spin-based magnetometers (e.g. NV centers) $\hat{Q}$ and $a$ correspond to the spin projection and magnetic field along a given direction, respectively, and $\lambda$ the gyromagnetic ratio \cite{RevModPhys.92.015004}. In this sense, our analysis based on Hamiltonian \eqref{eq:generic} is generic and readily extendable to a broad class of quantum accelerometers and sensors.\\
In estimation theory, the CFI quantifies how much information about an unknown parameter can be extracted from measurement data. It sets a lower bound on the variance of any unbiased estimator via the Cram\'er-Rao inequality \cite{Kay1993, LehmannCasella1998}. QFI generalizes this concept to quantum systems, defining the ultimate precision limit achievable in estimating parameters encoded into quantum states, accounting for quantum mechanical effects such as superposition and entanglement \cite{Braunstein1994, Giovannetti2011, Paris2009}. For pure states, the QFI determines the maximum achievable information over all possible measurements and is directly linked to the QCRB, which limits the variance in estimating $a$ to $1/(N\,H_Q)$, where $N$ is the number of experimental trials. In contrast, the CFI quantifies the sensitivity achievable by a specific measurement procedure, such as homodyne detection, and typically yields a lower bound that is greater than or equal to the QFI. When equality is achieved, the measurement is said to be optimal. Comparing the QFI and CFI thus provides a quantitative benchmark for how closely a given experimental scheme approaches the fundamental quantum limit. In what follows, we develop the general QFI formalism for the Hamiltonian~\eqref{eq:generic}.

During the interrogation time $\tau$, the probe evolves under the parameter-dependend Hamiltonian $\hat{H(a)}$, thereby encoding information about $a$ into its quantum state. This process is described by the unitary map
\begin{equation}
\label{eq:encoding}
\ket{\psi(a,\tau)}=U(a,\tau)\ket{\psi_0},\qquad
U(a,\tau)=\exp\!\Big[-\frac{i}{\hbar}\,\hat H(a)\,\tau\Big],
\end{equation}
with a normalized finite-energy probe $\ket{\psi_0}$. To quantify how uncertainty in the interrogation time affects the precision of estimating the parameter of interest $a$, we introduce the parameter vector $\theta=(a,\tau)$. Here, $\tau$ is treated as a nuisance parameter, i.e. a quantity that influences the probe dynamics, but is not itself estimated. The total differential of the quantum state is written as
\begin{equation}
\label{eq:totaldiff}
\ket{d\psi}=\partial_a\ket{\psi}\,da+\partial_\tau\ket{\psi}\,d\tau+O(\|d\theta\|^2) ,
\end{equation}
Physical states are represented by rays in Hilbert space: only variations of the state that are orthogonal to $\ket{\psi}$ correspond to observable changes. We therefore define the orthogonal projections of the tangents as
\begin{equation}
\label{eq:perptangent}
\ket{\partial_j\psi_\perp}:=(\mathbb I-\ket{\psi}\!\bra{\psi})\,\partial_j\ket{\psi},\qquad
\braket{\psi|\partial_j\psi_\perp}=0,\quad j\in\{a,\tau\} .
\end{equation}

We introduce the equivalent (Hermitian) local generators of parameter translation:
\begin{equation}
\label{eq:Ki_defs}
K_j:=i\,U^\dagger\partial_j U,\qquad
\tilde K_j:=i(\partial_j U)U^\dagger=U K_j U^\dagger,\qquad j\in\{a,\tau\}.
\end{equation}
They act in different pictures, but encode the same tangent. In terms of these generators, the state derivatives read
\begin{equation}
\label{eq:tangent_exact}
\partial_j\ket\psi=-\,i\,\tilde K_j\ket\psi,\qquad
\braket{\psi|\partial_i\psi}=-\,i\,\langle\tilde K_j\rangle,
\end{equation}
and the corresponding projected tangents become
\begin{equation}
\label{eq:proj_tangent_correct}
\ket{\partial_j\psi_\perp}=-\,i\big(\tilde K_j-\langle \tilde K_j\rangle\big)\ket{\psi},\qquad
\langle\cdot\rangle:=\bra{\psi}\cdot\ket{\psi}.
\end{equation}
For the time coordinate one has
\begin{equation}
\label{eq:Ktau_correct}
\partial_\tau U=-\tfrac{i}{\hbar}\hat H(a)\,U
\quad\Rightarrow\quad
\tilde K_\tau=\frac{\hat H(a)}{\hbar}
=K_\tau \, ,
\end{equation}
since $[\hat H(a),U(a,\tau)]=0$.
\\
The derivative $\partial_a U$, can be calculated via the Hadamard formula 
\begin{equation}
\label{eq:duhamel}
\partial_a e^{A(g)} = \int_0^1 e^{(1-s)A(g)}\,(\partial_g A(g))\,e^{sA(g)}\,ds \, .
\end{equation}
Setting $A(g) = -\frac{i}{\hbar} \, \hat{H}(a)\, \tau$ and $s' = s\,\tau$ yields
\begin{equation}
\partial_a U(a,\tau) = -\frac{i}{\hbar} \int_0^\tau e^{-\frac{i}{\hbar} H(a) (\tau - s')} \, (\partial_a H(a)) \, e^{-\frac{i}{\hbar} H(a) s'} \, ds' .
\end{equation}

Accordingly,
\begin{equation}
\label{eq:Kg}
K_a:= i\,U^\dagger\partial_a U
= \frac{1}{\hbar}\int_0^\tau \big(\partial_a H\big)_H(s)\,ds,
\qquad
(\cdot)_H(s):=e^{+\frac{i}{\hbar}H(a)s}(\cdot)\,e^{-\frac{i}{\hbar}H(a)s},
\end{equation}
and the state-acting generator is the unitarily conjugate operator
\begin{equation}
\tilde K_a:= i\,(\partial_a U)U^\dagger
= U\,K_a\,U^\dagger
= \frac{1}{\hbar}\int_0^\tau e^{-\frac{i}{\hbar}H(a)\,(\tau-s)}\;\partial_a H(a)\;e^{+\frac{i}{\hbar}H(a)\,(\tau-s)}\,ds \, .
\end{equation}
For the linear quantum sensor Hamiltonian \eqref{eq:generic}, $\partial_a H=\lambda\,Q$  gives
\begin{equation}
\label{eq:Kglinear}
K_a=\frac{\lambda}{\hbar}\int_0^\tau Q(s)\,ds \, .
\end{equation}

Normalization implies $\Re\,\braket{\psi|\partial_i\psi}=0$. Using \eqref{eq:perptangent} together with $\tilde K_j^\dagger=\tilde K_j$ and $\partial_j\ket\psi=-i\,\tilde K_j\ket\psi$,
\begin{equation}
\label{eq:projtangents}
\ket{\partial_j\psi_\perp}
= -\,i\big(\tilde K_j-\langle \tilde K_j\rangle\big)\ket{\psi},
\qquad
\langle\cdot\rangle:=\bra{\psi}\cdot\ket{\psi}.
\end{equation}
The Fubini-Study line element is
\begin{equation}
\label{eq:FSmetricdef}
ds^2=\sum_{i,j} g_{ij}\,d\theta_i\,d\theta_j , 
\end{equation}
where $g_{ij}:=\Re\braket{\partial_i\psi_\perp|\partial_j\psi_\perp}.$. A short calculation gives the covariance form
\begin{equation}
\label{eq:cov-sym}
g_{ij}
=\tfrac12\Big\langle\big\{\Delta\tilde K_i,\Delta\tilde K_j\big\}\Big\rangle
=\tfrac12\langle\{\tilde K_i,\tilde K_j\}\rangle-\langle \tilde K_i\rangle\langle \tilde K_j\rangle,
\qquad
\Delta\tilde K_j:=\tilde K_j-\langle\tilde K_j\rangle.
\end{equation}
For pure states the quantum Fisher information matrix equals four times the Fubini-Study metric, $F_{ij}=4\,g_{ij}$. Therefore, writing 
\begin{equation}
\label{eq:QFIM-pure}
ds^2=\tfrac14\Big[F_{aa}\,dg^2+2F_{a\tau}\,dg\,d\tau+F_{\tau\tau}\,d\tau^2\Big] ,
\end{equation}
and using Eq. \eqref{eq:Ktau_correct} and \eqref{eq:Kglinear}, we obtain
\begin{equation}
\label{eq:Fentries}
F_{aa}=4\,\Var(\tilde K_a),\qquad
F_{a\tau}=4\,\Cov\!\Big(\tilde K_a,\tfrac{H(a)}{\hbar}\Big),\qquad
F_{\tau\tau}=4\,\Var\!\Big(\tfrac{H(a)}{\hbar}\Big)=\frac{4(\Delta H)^2}{\hbar^2} \, .
\end{equation}
Positive semidefiniteness of $F$ yields the Cauchy-Schwarz constraint
\begin{equation}
\label{eq:CS}
F_{a\tau}^2\le F_{aa}\,F_{\tau\tau}.
\end{equation}

Treating $\tau$ as a nuisance parameter, we can eliminate $d\tau$ by minimizing the infinitesimal Bures distance $ds^2$ at fixed $da$:
\begin{equation}
\label{eq:opt-dtau}
\partial_{(d\tau)}\,ds^2=\tfrac12\big(F_{a\tau}\,da+F_{\tau\tau}\,d\tau\big)=0
\ \Rightarrow\
d\tau^\star=-\frac{F_{a\tau}}{F_{\tau\tau}}\,da\qquad (F_{\tau\tau}>0).
\end{equation}
Substituting this back into $ds^2$ gives $ds_{\min}^2=(1/4) F_a^{\mathrm{eff}}$, where 
\begin{equation}
\label{eq:Feff-geom}
F_a^{\mathrm{eff}}:=F_{aa}-\frac{F_{a\tau}^2}{F_{\tau\tau}} ,
\end{equation}
is the effective Fisher information for the parameter $a$, corresponding to the Schur complement of the nuisance block in the full Fisher information matrix. If $F_{\tau\tau}=0$, Eq. \eqref{eq:Feff-geom} should be interpreted using the Moore-Penrose pseudoinverse of $F$, which reproduces the same Schur complement on the estimable subspace. {Notice that Eq. \eqref{eq:Feff-geom} admits a direct operational interpretation: the second term subtracts the component of the signal generator that is statistically indistinguishable from a fluctuation in time. In this sense, the Schur complement removes the information about $a$ that is lost due to its correlation with time uncertainty, leaving only the information that can be accessed independently of the nuisance parameter.}

For $N$ independent probes this yields the QCRB, with time treated as a nuisance parameter
\begin{equation}
\label{eq:QCRB-nuis}
\Var(\hat a)\ \ge\ \frac{1}{N\,F_a^{\mathrm{eff}}}
=\frac{1}{N}\,(F^{-1})_{aa}
=\frac{1}{N\,F_{aa}}\cdot\frac{1}{1-\rho^2} , 
\end{equation}
where \( \rho^2 = F_{a\tau}^2/(F_{aa} F_{\tau \tau}) \) is the squared correlation coefficient between the estimators of \( a \) and \( \tau \). Since $0 \le \rho^2 \le 1$, the two-parameter bound (\eqref{eq:QCRB-nuis}) is always larger than (or equal to) the standard single-parameter limit. This degradation can be quantified by the inflation factor 
\begin{equation} 
\varepsilon = \frac{1}{1 - \rho^2} ,
\end{equation}
which measures the strength of the coupling between \( a \) and \( \tau \). The closer $\rho^2$ is to unity, the stronger the parameter correlation and the greater the degradation of precision in estimating $a$.

We observe that in the full quantum estimation formalism, projecting the partial Symmetric Logarithmic Derivative (SLD) onto the orthogonal subspace of the nuisance parameter yields the same effective Fisher information (Schur complement) \cite{Suzuki2020Nuisance}.
Let $L_i$ denote the SLDs,
\begin{equation}
\label{eq:SLD}
\partial_i\rho=\tfrac12(L_i\rho+\rho L_i),\qquad
F_{ij}=\tfrac12\,\Tr\big[\rho\{L_i,L_j\}\big].
\end{equation}
Defining the SLD inner product $\langle\!\langle A,B\rangle\!\rangle_\rho:=\tfrac12\Tr[\rho\{A,B\}]$, the efficient (nuisance-orthogonal) score for the parameter of interest is constructed by projecting $L_a$ onto the subspace orthogonal to the nuisance SLD $L_\tau$:
\begin{equation}
\label{eq:efficient}
\widetilde L_a:=L_a-\Pi(L_a\,|\,L_\tau),\qquad
\Pi(L_a\,|\,L_\tau)=\frac{\langle\!\langle L_a,L_\tau\rangle\!\rangle_\rho}{\langle\!\langle L_\tau,L_\tau\rangle\!\rangle_\rho}\,L_\tau
=\frac{F_{a\tau}}{F_{\tau\tau}}\,L_\tau\quad(F_{\tau\tau}>0),
\end{equation}
for which $\langle\!\langle \widetilde L_a,L_\tau\rangle\!\rangle_\rho=0$. The corresponding effective Fisher information is
\begin{equation}
\label{eq:Feff-SLD}
F_a^{\mathrm{eff}}=\langle\!\langle \widetilde L_a,\widetilde L_a\rangle\!\rangle_\rho
=F_{aa}-\frac{F_{a\tau}^2}{F_{\tau\tau}},
\end{equation}
in agreement with \eqref{eq:Feff-geom}. This shows that \eqref{eq:QCRB-nuis} correctly captures the quantum information for $a$ in the presence of time uncertainty. Moreover, attainability of the above limit is guaranteed: Theorem 5.3 of \cite{Suzuki2020Nuisance} establishes that the minimum locally unbiased mean-square error for the parameter of interest equals the inverse of the partial SLD Fisher information (i.e.\ the Schur complement above) and that there exists an optimal measurement and estimator achieving this bound. The SLD-compatibility (weak-commutativity) condition concerns the simultaneous attainability of the full multi-parameter QCRB; it is not required when optimizing precision for a single parameter in the presence of nuisance parameters (see Sec. III of Ref.\cite{SidhuKok2020Multiparameter}). Therefore, the quantum limit \eqref{eq:QCRB-nuis} is asymptotically attainable without imposing $[L_a,L_\tau]=0$.

We now examine how this limit connects to the fundamental bounds on quantum evolution. Along the physical-time axis (using $\tilde K_\tau = H/\hbar$), the statistical speed is
\begin{equation}
\label{eq:MTspeed}
\frac{ds}{d\tau}=\sqrt{g_{\tau\tau}}
=\sqrt{\Var\!\Big(\tfrac{H}{\hbar}\Big)}
=\frac{\Delta H}{\hbar}.
\end{equation}
For $N$ independent repetitions of the same quantum state, the quantum Cram\'er-Rao bound for estimating $\tau$ reads
\begin{equation}
\label{eq:QCRB-tau}
\Delta\tau\;\ge\;\frac{1}{\sqrt{N\,F_{\tau\tau}}}
=\frac{\hbar}{2\sqrt{N}\,\Delta H},
\end{equation}
where $F_{\tau\tau}=4\,\Var(H/\hbar)$. 
The corresponding single-probe limit ($N=1$) gives $\Delta\tau\ge 1/\!\sqrt{F_{\tau\tau}}$, which coincides exactly with the Mandelstam-Tamm bound on the minimal timescale of quantum evolution. The factor $1/\sqrt{N}$ reflects the improved statistical accuracy from averaging over $N$ independent probes. 
Since $\Var(\hat a)$  depends explicitly on the Mandelstam-Tamm scale through the term $F_{\tau\tau}$, the degradation of sensitivity in estimating $a$ arises from this fundamental quantum speed limit, and not from an external timing error or lack of synchronization with a classical clock. 

We finally observe that, at special “decoupling” interrogation times,
\begin{equation}
\label{eq:decouple}
\Cov\!\Big(\int_0^\tau Q(s)\,ds,\;H(a)\Big)=0,
\end{equation}
one has $F_{a\tau}=0$ and the inflation factor is $\varepsilon=1$.
This means that in an idealized, instantaneous measurement picture, probing the system exactly at those times would be equivalent to measuring $a$ without any influence from uncertainty in $\tau$. In practice, however, any realistic sensing protocol averages the signal over a finite time interval. Therefore, the single-parameter limit cannot be exactly attained as a sensitivity bound. In a continuous measurement scheme, even if $F_{a\tau}$ is not zero for all times, one could in principle tune system parameters such that its time average over a full measurement cycle is zero. In this case, the effective QFI matrix for a continuous measurement with stationary statistics (such as homodyne detection) becomes block-diagonal on average, and the effective QCRB approximately reproduces the single-parameter limit. It is important to remark, however, that instantaneous coupling between the parameters may still be large and that a vanishing time-average in general cannot eliminate transient correlations or conditioning issues in the inversion of the integrated Fisher matrix. We will further discuss these aspects in the specific case of our optomechanical gravimeter.

 
\section{Optomechanical gravimeter: QFI and QCRB}\label{section_iii}

The general results of Sec. II provide the fundamental benchmark against which any specific quantum sensor \eqref{eq:generic}, together with its associated sensing protocol, can be evaluated.
    
As an illustrative case, we consider the cavity-optomechanical gravimeter investigated in Ref. \cite{NonlinearGravimetry2018,qvarfort2021optimal}. For this system, homodyne detection provides a natural measurement strategy whose CFI can be compared directly with the QFI to assess optimality. This comparison is developed in the following sections.

 The Hamiltonian of the system reads
	\begin{equation}
	\hat H_{g}=\hbar\omega_{c}\hat a^{\dagger}\hat a
	+\hbar\omega_{m}\hat b^{\dagger}\hat b
	-\hbar k\,\hat a^{\dagger}\hat a(\hat b^{\dagger}+\hat b)
	+m g\cos\theta\,\hat x_{o} \, .
	\label{eq:HamGrav}
	\end{equation}
	
This Hamiltonian provides a particularly clear example of the generic model (\ref{eq:generic}). The free part  
	$\hat H_{0}=\hbar\omega_{c}\hat a^{\dagger}\hat a+\hbar\omega_{m}\hat b^{\dagger}\hat b -\,\hbar k\,\hat a^{\dagger}\hat a(\hat b^{\dagger}+\hat b)$
contains the photon number \(\hat a^{\dagger}\hat a\) of a single cavity mode at frequency \(\omega_{c}\) and the phonon number \(\hat b^{\dagger}\hat b\) of a mechanical resonator at \(\omega_{m}\). The resonator position
	\(\hat Q=\hat x_{o}=\sqrt{\hbar/(2m\omega_{m})}\,(\hat b^{\dagger}+\hat b)\) acts the readout variable, while the cavity photon number serves as the probe via the optomechanical coupling term \(\hbar k\,\hat a^{\dagger}\hat a(\hat b^{\dagger}+\hat b)\). The coefficient $\lambda=m\cos\theta$ quantifies the strength of the coupling to the gravitational potential, where $m$ is the mass of the mechanical oscillator and $\theta$ is the angle between the oscillator's displacement direction and the vertical axis.
   
	We now evaluate the QFI to determine the sensitivity limits of the system (\ref{eq:HamGrav}) for estimating $g$ in the presence of time-uncertainty, and compute the corresponding CFI for homodyne detection to assess the performance of this measurement strategy.
	We focus on the case where the system is prepared in a pure state. In the rotating frame, such state can be written as 
	\begin{equation}
	\label{state}
	\ket{\Psi(\bar{g},t)} = \sum_{n=0}^{\infty} c_{n}(\bar{g},t)\,\ket{n}_{C}\otimes \ket{\gamma_{n}(\bar{g},t)}_{O}\,.
	\end{equation}
	where $\bar{g} = A\,g$ is the scaled coupling parameter, with
	$A = \cos\theta\,\sqrt{m/(2\hbar\omega_m^3)}$ and the dimensionless time $t=\omega_m t_{phys}$, where $t_{phys}$ is the physical time.
	Here, \(n\) labels the Fock states of the cavity mode (indicated by the subscript \(C\)), while \(\ket{\gamma_{n}(\bar{g},t)}_{O}\) denotes the oscillator (or displacement) state (subscript \(O\)). This representation clearly shows that the overall state is a superposition of products of cavity and oscillator states, with the coefficients \(c_n(\bar{g},t)\) determining the weight and phase of each component.
	
	The coefficients \(c_n(\bar{g},t)\) are given by
	\begin{equation}
	\label{cn}
	\begin{aligned}
	c_n(\bar{g},t)
	&=
	e^{-|\alpha|^2/2}\,\frac{\alpha^n}{\sqrt{n!}}
	\exp\Biggl\{
	i\Bigl[\bar{k}^2\,n^2-2\,\bar{k}\,\bar{g}\,n\Bigr](t-\sin t)
	+\frac{1}{2}\Bigl[(\bar{k}n-\bar{g})(\eta^*\beta-\eta\beta^*)\Bigr]
	\Biggr\},\\[1mm]
	|c_n|^2
	&=
	\left|P_n(\alpha)\right|^2\,\exp\Biggl\{2\,\mathrm{Re}\Bigl[
	i\Bigl(\bar{k}^2n^2-2\,\bar{k}\,\bar{g}\,n\Bigr)(t-\sin t)
	+\frac{1}{2}(\bar{k}n-\bar{g})(\eta^*\beta-\eta\beta^*)
	\Bigr]\Biggr\}\,,
	\end{aligned}
	\end{equation}
	\\
	\\
    The factor 
	\(
	P_n(\alpha)=e^{-|\alpha|^2/2}\,\frac{\alpha^n}{\sqrt{n!}},
	\)
	represents the standard coherent state expansion, and the additional exponential factor introduces a phase that depends nonlinearly on \(n\) as well as on the parameters \(\bar{g}\) and \(t\). 
Following Ref.~\cite{NonlinearGravimetry2018}, we assume that also the oscillator is initialized in a coherent state of amplitude $\beta$ and we define the normalized optomechanical coupling as $\bar{k} = k/\omega_m$. 
	The time-dependent quantities
	\begin{equation}
	\zeta(t) = t - \sin t, \quad \eta = 1 - e^{-it}, \quad \eta^* = 1 - e^{it},
	\end{equation}
encode essential aspects of the system's temporal evolution. In particular, the phase factor $\zeta(t)$ reflects the nonlinear temporal behavior of the mechanical oscillator. This nonlinearity, intrinsic to the optomechanical interaction, leads to a nontrivial accumulation of phase over time, which significantly influences the evolution of the system.
	
Next, the oscillator states, which depend on the same parameters, are given by
	\begin{equation}
	\label{gamman}
	\begin{aligned}
	\ket{\gamma_n(\bar{g},t)}_{O} &= \ket{\gamma_n(\bar{g},t)},\\[1mm]
	\gamma_n(\bar{g},t)
	&=
	e^{-it}\,\beta+(\bar{k}\,n-\bar{g})\Bigl[1-e^{-it}\Bigr]\,.
	\end{aligned}
	\end{equation}
This equation shows that the displacement \(\gamma_n(\bar{g},t)\) is a linear function of \(n\) and \(\bar{g}\), and is modulated by the time-dependent factors \(e^{-it}\) and \(1-e^{-it}\). It is the interplay between these terms that ultimately leads to the nontrivial correlations between the optical and mechanical parts of the state.
	
We now calculate QFI matrix coefficients for the system \eqref{eq:HamGrav}. For a pure state \(\ket{\Psi(\theta_{1},\theta_{2})}\) that depends on two parameters \(\theta_{1}\) and \(\theta_{2}\), the general expression for the QFI matrix element is given by
	\begin{equation}
	\label{QFI_general}
	\begin{aligned}
	F_{\theta_1 \, \theta_2} 
	=\; 4\,\mathrm{Re}\Biggl\{
	\sum_{n=0}^\infty \Bigl[
	&(\partial_{\theta_1} c_n)^*(\partial_{\theta_2} c_n)
	+ (\partial_{\theta_1} c_n)^*\,c_n\,\bra{\gamma_n}\partial_{\theta_2}\ket{\gamma_n}\\[1mm]
	&+ c_n^*(\partial_{\theta_2} c_n)\,\bra{\partial_{\theta_1}\gamma_n}\gamma_n\rangle
	+ |c_n|^2\,\bra{\partial_{\theta_1}\gamma_n}\partial_{\theta_2}\ket{\gamma_n}
	\Bigr]\\[1mm]
	&-\Biggl(
	\sum_{n=0}^\infty \Bigl[(\partial_{\theta_1} c_n)^*\,c_n+|c_n|^2\,\bra{\partial_{\theta_1}\gamma_n}\gamma_n\rangle\Bigr]
	\Biggr)
	\\[1mm]
	&\quad\times\;
	\Biggl(
	\sum_{m=0}^\infty \Bigl[c_m^*(\partial_{\theta_2} c_m)+|c_m|^2\,\bra{\gamma_m}\partial_{\theta_2}\ket{\gamma_m}\Bigr]
	\Biggr)
	\Biggr\}\, ,
	\end{aligned}
	\end{equation}
where, in our case, \(g\) is the parameter of interest and \(t\) is a nuisance parameter, i.e. a quantity that influences the estimation of $g$, but is not itself estimated. Eq. \eqref{QFI_general} involves derivatives of both the coefficients and the oscillator states and includes extra terms (known as “score terms”) to ensure invariance under local phase changes.
	
To proceed, we need to compute the derivatives of the coefficient functions. We set \(\theta_1 \to g\) and \(\theta_2 \to t\)) and define
	\(
	\partial_g c_n \equiv (g\,c_n),\quad \partial_t c_n \equiv (t\,c_n)\,.
	\)
	A straightforward calculation shows that
\begin{align}\label{cf_eq}
\partial_{\bar{g}} c_n &= c_n\,F_{\bar{g}}(n), &
F_{\bar{g}}(n) &= -\,2i\,\bar{k}\,n\,\zeta \;-\; \tfrac12\big(\eta^\ast \beta - \eta \beta^\ast\big),\nonumber\\[1mm]
\partial_{t} c_n &= c_n\,F_{t}(n), &
F_{t}(n) &= i\big(\bar{k}^2 n^2 - 2\bar{k}\,\bar{g}\,n\big)\,(1-\cos t)
\;-\; \tfrac{i}{2}\,X\big(\beta e^{it}+\beta^\ast e^{-it}\big) \, ,
\end{align}

where $X=\bar k n-\bar g$. The functions act as local generators for variations in the parameter \(g\) and \(t\) respectively, and describe how the state changes in response to infinitesimal changes in these parameters.
	
The derivatives of the oscillator states are defined as
	$\partial_t\ket{\gamma_n} \equiv \ket{t\,\gamma_n},\quad \partial_g\ket{\gamma_n} \equiv \ket{g\,\gamma_n}\,.$
	The inner products between these derivatives and the original oscillator states are crucial in the evaluation of the QFI. For example, one finds
	\begin{align}\label{gg_eq}
\langle \gamma_n \mid \bar g\,\gamma_n \rangle
&=\tfrac12\big(\eta^*\gamma_n-\eta\,\gamma_n^*\big),\nonumber\\
\langle \bar g\,\gamma_n \mid \gamma_n \rangle
&=-\,\tfrac12\big(\eta^*\gamma_n-\eta\,\gamma_n^*\big),\nonumber\\
\langle \bar g\,\gamma_n \mid \bar g\,\gamma_n \rangle
&=\tfrac12|\eta|^2+\tfrac14\big|\eta^*\gamma_n-\eta\,\gamma_n^*\big|^2.
\end{align}
	These expressions quantify the overlap between the oscillator state and its derivative with respect to \(g\), and they will later appear in the expression for \(F_{g,g}\).
	
	Similarly, the overlaps involving the time derivatives are given by
	\begin{align}\label{tt_eq}
\langle \gamma_n \mid t\,\gamma_n \rangle
&=\tfrac12\big(\gamma_n^* X_n-\gamma_n X_n^*\big),\nonumber\\
\langle t\,\gamma_n \mid \gamma_n \rangle
&=-\,\tfrac12\big(\gamma_n^* X_n-\gamma_n X_n^*\big),\nonumber\\
\langle t\,\gamma_n \mid t\,\gamma_n \rangle
&=\tfrac12|X_n|^2+\tfrac14\big|\gamma_n^* X_n-\gamma_n X_n^*\big|^2.
\end{align}

	where we introduce the variable
	\begin{equation}
	X_n = i\,e^{-it}\Bigl[(\bar{k}n-g)-\beta\Bigr]\,.
	\end{equation}
	to simplify the notation. These inner products encode the sensitivity of the oscillator state with respect to changes in the time parameter.
	
	In addition, a mixed derivative overlap is given by
	\begin{align}\label{gt_eq}
\langle \bar g\,\gamma_n \mid t\,\gamma_n \rangle
&=\tfrac12(-\eta^*)\,X_n
+\tfrac14\big(\eta^*\gamma_n-\eta\,\gamma_n^*\big)\big(\gamma_n^* X_n-\gamma_n X_n^*\big).
\end{align}
	This term will appear in the off-diagonal component of the QFI matrix and represents the coupling between the two parameters \(g\) and \(t\).
	
	We now present the final compact expressions for the QFI components. The \(g\)-component of the QFI is given by
	\begin{equation}
	\label{Fgg_final}
	\begin{aligned}
	F_{gg}
	=\; 4\,\mathrm{Re}\Biggl\{
	\sum_{n=0}^\infty |c_n|^2\,\Bigl[
	&|F_g(n)|^2
	+ F_g(n)^*\bra{\gamma_n}\ket{g\,\gamma_n}\\[1mm]
	&+ F_g(n)\bra{g\,\gamma_n}\ket{\gamma_n}
	+ \bra{g\,\gamma_n}\ket{g\,\gamma_n}
	\Bigr]\\[1mm]
	&-\Biggl(
	\sum_{n=0}^\infty |c_n|^2\,\Bigl(F_g(n)^*+\bra{g\,\gamma_n}\ket{\gamma_n}\Bigr)
	\Biggr)^2
	\Biggr\}\,.
	\end{aligned}
	\end{equation}
	This equation combines contributions from both the derivative of the coefficients and the overlaps of the oscillator state derivatives with the state itself. The term \(F_g(n)\) encodes the sensitivity of the state with respect to the parameter \(g\).
	
	Similarly, the QFI element corresponding to the time parameter \(t\) is given by
	\begin{equation}
	\label{Fttexpression}
	\begin{aligned}
	F_{tt}
	=\; 4\,\mathrm{Re}\Biggl\{
	\sum_{n=0}^\infty \Bigl[
	&(t\,c_n)^*(t\,c_n)
	+(t\,c_n)^*\,c_n\,\bra{\gamma_n}\ket{t\,\gamma_n}\\[1mm]
	&+\, c_n^*(t\,c_n)\,\bra{t\,\gamma_n}\ket{\gamma_n}
	+\, |c_n|^2\,\bra{t\,\gamma_n}\ket{t\,\gamma_n}
	\Bigr]\\[1mm]
	&-\Biggl(
	\sum_{n=0}^\infty \Bigl[(t\,c_n)^*\,c_n+|c_n|^2\,\bra{t\,\gamma_n}\ket{\gamma_n}\Bigr]
	\Biggr)^2
	\Biggr\}\,.
	\end{aligned}
	\end{equation}
	
	By using the relation $(t\,c_n)= c_n\,F_t(n)$, we can rewrite the above expression as
	\begin{equation}
	\label{FttFinal}
	\begin{aligned}
	F_{tt}
	=\; 4\,\mathrm{Re}\Biggl\{
	\sum_{n=0}^\infty |c_n|^2\Bigl[
	&|,F_t(n)|^2
	+,F_t(n)^*\bra{\gamma_n}\ket{t\,\gamma_n}\\[1mm]
	&+,F_t(n)\bra{t\,\gamma_n}\ket{\gamma_n}
	+\bra{t\,\gamma_n}\ket{t\,\gamma_n}
	\Bigr]\\[1mm]
	&-\Biggl(
	\sum_{n=0}^\infty |c_n|^2\,\Bigl(F_t(n)^*+\bra{t\,\gamma_n}\ket{\gamma_n}\Bigr)
	\Biggr)^2
	\Biggr\}\,.
	\end{aligned}
	\end{equation}
	
	The off-diagonal component, which represents the coupling between the parameters \(g\) and \(t\), is given by
	\begin{equation}
	\label{Ft_gExpression}
	\begin{aligned}
	F_{tg}
	=\; 4\,\mathrm{Re}\Biggl\{
	\sum_{n=0}^\infty \Bigl[
	&(t\,c_n)^*(g\,c_n)
	+(t\,c_n)^*\,c_n\,\bra{\gamma_n}\ket{g\,\gamma_n}\\[1mm]
	&+\, c_n^*(g\,c_n)\,\bra{t\,\gamma_n}\ket{\gamma_n}
	+\, |c_n|^2\,\bra{t\,\gamma_n}\ket{g\,\gamma_n}
	\Bigr]\\[1mm]
	&-\Biggl[
	\sum_{n=0}^\infty \Bigl((t\,c_n)^*\,c_n+|c_n|^2\,\bra{t\,\gamma_n}\ket{\gamma_n}\Bigr)
	\Biggr]\\[1mm]
	&\quad\times
	\Biggl[
	\sum_{m=0}^\infty \Bigl(c_m^*(g\,c_m)+|c_m|^2\,\bra{\gamma_m}\ket{g\,\gamma_m}\Bigr)
	\Biggr]
	\Biggr\}\,.
	\end{aligned}
	\end{equation}
	After substitution of Eqs. \eqref{cf_eq}, \eqref{gg_eq}, \eqref{tt_eq} and \eqref{gt_eq} in the QFI matrix elements \eqref{Fgg_final},\eqref{FttFinal}, and \eqref{Ft_gExpression}, and using the Poisson moments for the photon-number distribution $\langle n\rangle=\mu$, $\mathrm{Var}(n)=\mu$, together with the relations $\langle X\rangle=\bar k\mu-\bar g$, $\mathrm{Var}(X)=\bar k^2\mu$ and $\mathrm{Cov}(n,X)=\bar k\mu$ we obtain the explicit expressions

\begin{align}
\label{eq:Fgg-g-theta0}
F_{gg}(t)
&=\Big(\frac{m}{2\hbar\,\omega_m^{3}}\Big)\!\left[\,16\,\bar k^2\mu\,\zeta^2\ +\ 4\,(1-\cos t)\,\right],\\[2mm]
\label{eq:Ftt-g-theta0}
F_{tt}(t)
&=\ 2\Big[\bar k^2\mu+\big(\bar k\mu-\sqrt{\tfrac{m}{2\hbar\,\omega_m^{3}}}\,g-\beta_R\big)^2+\beta_I^2\Big],\\[2mm]
\label{eq:Fgt-g-theta0}
F_{g t}(t)
&=\sqrt{\tfrac{m}{2\hbar\,\omega_m^{3}}}\,
\Big[-16\,\bar k^2\mu\,\zeta(1-\cos t)\,\beta_R\ +\ 2\big(-\sin t\,(\bar k\mu-\sqrt{\tfrac{m}{2\hbar\,\omega_m^{3}}}\,g-\beta_R)+(1-\cos t)\beta_I\big)\Big] \, ,
\end{align}
where $\beta_R\equiv \Re \beta$ and $\beta_I\equiv \Im \beta$.	
	
\begin{figure}
	    \centering
	    \includegraphics[width=1\linewidth]{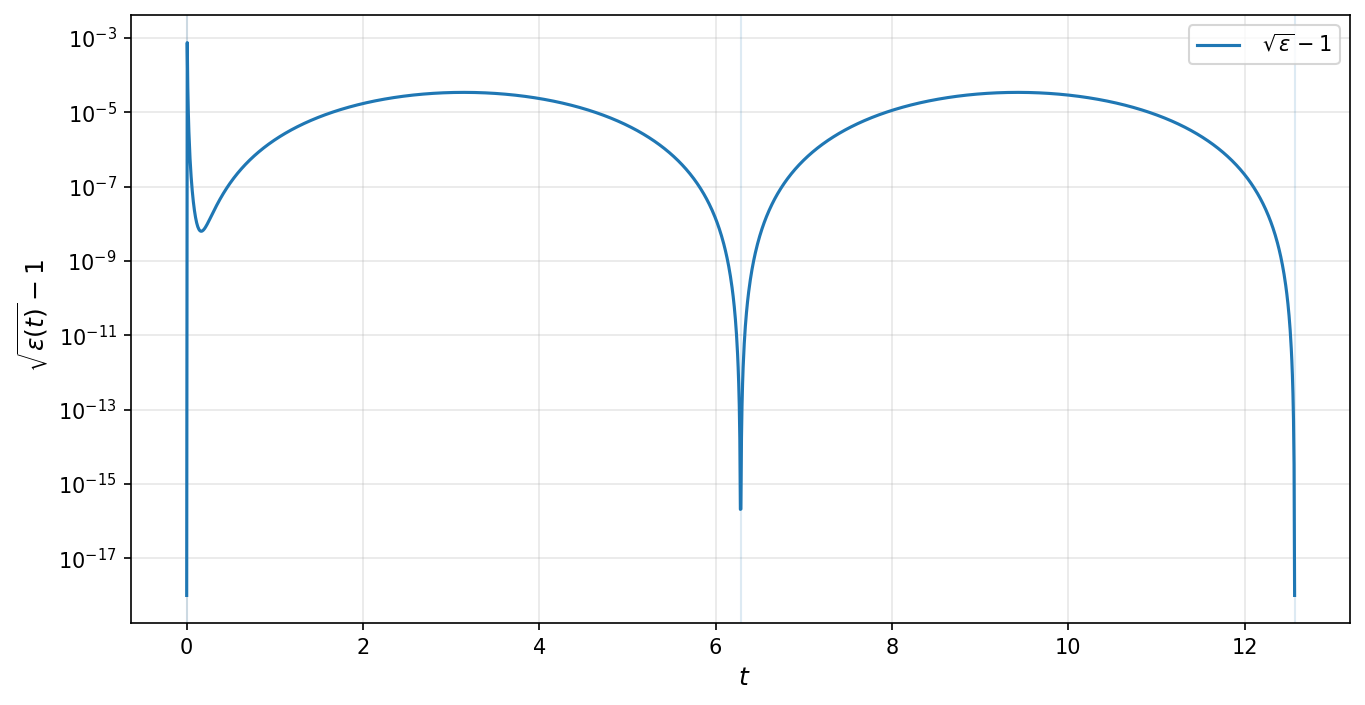}
	     \caption{Relative degradation in sensitivity caused by the correlation with the nuisance parameter $t$: $\sqrt{\varepsilon(t)}-1=\Delta g (t)/\Delta g_{\mathrm{single}}(t)-1$. Optomechanical parameters: $m=10^{-14}\,\mathrm{kg}$, $\omega_m=100\,\mathrm{rad\,s^{-1}}$, $\bar k=1963$, $\mu=10^{6}$, $\beta_R=1$, $\beta_I=0$ (cf Ref. \cite{NonlinearGravimetry2018}).
         }
  \label{fig1}
	\end{figure}

As discussed in Sec. II, when $F_{tg} = 0$ the estimation of the parameter of interest decouple from time uncertainty and the QCRB reduces to the standard limit $\Delta g_{\mathrm{single}}(t) \ge 1/F_{gg}$, where $F_{gg}$ here is the single-parameter QFI component with respect to $g$. This occurs at stroboscopic times $t=2\pi\ell$ ($\ell\in\mathbb{Z}$) where one has
\begin{align}
F_{gg}(2\pi\ell)
&=\Big(\frac{m}{2\hbar\,\omega_m^{3}}\Big)\,16\,\bar k^2\mu\,\zeta^2
=\frac{32\,\pi^2\,m}{\hbar\,\omega_m^{3}}\;\bar k^2\,\mu\,\ell^2,\\[1mm]
F_{tt}(2\pi\ell)
&=2\Big[\bar k^2\mu+\big(\bar k\mu-A g-\beta_R\big)^2+\beta_I^2\Big],
\qquad
A=\sqrt{\frac{m}{2\hbar\,\omega_m^{3}}},\\[1mm]
F_{gt}(2\pi\ell)
&=0.
\end{align} \, 
and, consequently, 
\begin{equation}
\mathrm{Var}(g)\ \ge\ \frac{1}{F_{gg}(2\pi\ell)}
=\frac{\hbar\,\omega_m^{3}}{32\,\pi^2\,m\,\bar k^2\,\mu\,\ell^2}\,.
\label{IL}
\end{equation}
In Fig.\ref{fig1}, we plot the relative difference $\sqrt{\varepsilon(t)}-1$ between the QCRB obtained treating time as a nuisance parameter and the single-parameter limit $\Delta g_{\mathrm{single}}(t)$, for the optomechanical parameters reported in  Ref. \cite{NonlinearGravimetry2018} (see caption). The deviation between the two limits shows an explicit time dependence with maxima around values of $10^{-5}$ and minima at stroboscopic times, where the single-parameter bound and the one with time treated as a nuisance parameter coincide (within the numerical precision of the plot).
At these times, the optical cavity state is completely decoupled from the oscillator, meaning that all information about $g$ is transferred to the phase of the cavity state  \cite{NonlinearGravimetry2018}. As a result, in an ideal measurement scheme, it would be sufficient to probe only the cavity state at the end of one oscillation period, without any need to access the mechanical oscillator state. On the other hand, decoherence arising from damping to the oscillator motion during the state evolution will adversely affect the final measurement sensitivity and cause the oscillator state to evolve in a mixed state. Following Ref.\cite{NonlinearGravimetry2018}, we assume that the mechanical element remains coherent over one full oscillation period, and focus only on the time $t=2 \pi$. At this time, the sensitivity is equal to $\Delta g(2\pi)\approx 2.94\times10^{-15}\,\mathrm{m\,s^{-2}}$ for the above optomechanical parameters.

As already remarked in Sec. \ref{section_ii}, the single-parameter limit should be interpreted as an ideal bound, achievable only in case of perfectly instantaneous measurements. In practice, any measurement scheme (e.g. homodyne detection, see Sec. \ref{section_IV}) produces a time series over a finite duration. The effective two-parameter Fisher information is integrated over the measurement interval and the averaged off-diagonal term $F_{gt}$ is always finite, although its contribution is minimized near stroboscopic times.

On the other hand, one could, at least in principle, engineer the system such that $\langle F_{gt}\rangle $=0, where here $\langle \, \cdot \, \rangle$ denotes a time average. In this case, the coupling between the parameter of interest and the nuisance variable cancels on average in a continuous time measurement, thus recovering an approximate Heisenberg-limited sensitivity. Considering a measurement of duration $T$, where $T$ is large with respect to the oscillator period ($T \gg 1$), the time average of \eqref{eq:Fgt-g-theta0} gives the decoupling condition 
\begin{equation}
\beta_I = {4 \bar{k}^2 \mu \beta_R T}
\label{decc}
\end{equation}
For a mechanical oscillator cooled down to its quantum ground state ($\beta = 0$), this condition is always satisfied. In this case, the effects of time uncertainty on the estimation precision of $g$ are averaged out in continuous-time measurements. It is important to remark, however, that instantaneous coupling between the parameters may still be large ($F_{gt} \neq 0$, see Eq. (47)) and that a vanishing time-average, in general, cannot eliminate transient correlations.
More generally, the condition \eqref{decc} could be approached by preparing the mechanical coherent state with the appropriate phase (via calibrated resonant, impulsive drives), setting the intracavity photon number with the input power, and choosing the measurement duration. While implementing such a complex control strategy would not significantly improve sensitivity (as we have seen, for typical optomechanical paramters the relative deviation between the two limits remains bounded at approximately $10^{-5}$) it could nevertheless provide a means to directly probe the effects of time uncertainty in future quantum-limited experiments. \\

\section{Homodyne detection: CFI Matrix and Cramér-Rao Bound} \label{section_IV}

We now calculate the CFI for homodyne detection of the optical field nd analyze the conditions under which this measurement saturates the QCRB derived earlier.
In our optomechanical system, a homodyne measurement of a quadrature \(x_\lambda\) produces a probability density distribution of outcomes
    \begin{equation}\label{eq:p_x_final}
    p(x_\lambda;g,t) \;=\; e^{-|\alpha|^2}\,\sum_{n,n'} \mathcal{C}_{n,n'}(g,t)\,\mathcal{F}_{n,n'}(g,t)\,d_{n,n'}(x_\lambda) , 
    \end{equation}
    which encodes the interplay between the optical and mechanical degrees of freedom. The parameter of interest $g$ and the nuisance parameter $t$ enter through the optical phase and the mechanical displacement.
    The factor \(e^{-|\alpha|^2}\) represents the Poissonian weight of the coherent state, and the sum over indices \(n\) and \(n'\) accounts for contributions from different Fock states. The functions \(\mathcal{C}_{n,n'}(g,t)\), \(\mathcal{F}_{n,n'}(g,t)\), and \(d_{n,n'}(x_\lambda)\) respectively characterize the phase evolution, the mechanical state overlap, and the detection response.
    
    The coefficients \(\mathcal{C}_{n,n'}(g,t)\) capture the main phase dependence of the optical field and are given by 
    \begin{equation}
    \mathcal{C}_{n,n'}(g,t) \;=\; \frac{\alpha^n (\alpha^*)^{n'}}{\sqrt{n!\,n'!}} \exp\!\Bigl\{ i\Bigl[\bar{k}^2 (n^2-n'^2)-2\bar{k}\,\bar{g}(n-n')\Bigr]\zeta(t) \Bigr\}.
    \end{equation}
This expression captures the contribution of each cavity Fock state to the overall phase evolution. The Poissonian pre-factor reflects the coherent-state weights, while the exponential term contains the nonlinear optomechanical phase shifts.

The function 
  {  \begin{equation}
   \mathcal{F}_{n,n'}(g,t)\;=\;\exp\!\Bigl\{\frac{1}{2}\bar{k}(n-n')\Bigl(\eta^*\,\beta-\eta\,\beta^*\Bigr)
\;-\;\frac{1}{2}\,|\gamma_n|^2\;-\;\frac{1}{2}\,|\gamma_{n'}|^2\;+\;\gamma_{n'}^*\gamma_n\Bigr\}.
,
    \end{equation}}
    quantifies the overlap between different mechanical coherent states. The auxiliary variable $\gamma_n$, introduced in (\ref{gamman}), captures the displacement of the mechanical oscillator, and together with the previously introduced variables $\eta$, $\eta^*$, links the optomechanical dynamics to the observed quadrature fluctuations.

    The detection response terms $d_{n,n'}(x_\lambda)$ can be written as \cite{qvarfort2021optimal}
    \begin{equation}
        d_{n,n'}(x_\lambda)=\frac{e^{-x_\lambda^{2}}}{\pi^{1/2}}\,
\frac{H_{n}(x_\lambda)\,H_{n'}(x_\lambda)\,e^{-i\lambda (n-n')}}
     {2^{(n+n')/2}\,\sqrt{n!\,n'!}}
    \end{equation}
    These functions \(d_{n,n'}(x_\lambda)\) encode the homodyne detection response by representing the overlaps of the Fock states with the quadrature eigenstate \(\ket{x_\lambda}\). This connection ensures that the theoretical model properly reflects the experimental measurement process.
    
    The CFI for a parameter \(\theta\) is defined by 
    \begin{equation}\label{eq:Fisher_def_general}
    \mathcal{F}_{\theta\theta} \;=\; \int dx_\lambda\, \frac{\bigl[\partial_\theta p(x_\lambda;g,t)\bigr]^2}{p(x_\lambda;g,t)}.
    \end{equation}
    This definition quantifies the sensitivity of the probability distribution to changes in the parameter \(\theta\) and sets a lower bound on the variance of any unbiased estimator of \(\theta\).
    
    When estimating more than one parameter simultaneously, the full CFI matrix is constructed as 
    \begin{equation}
    \mathbf{\mathcal{F}} \;=\;
    \begin{pmatrix}
    \mathcal{F}_{gg} & \mathcal{F}_{gt}\\[1mm]
    \mathcal{F}_{tg} & \mathcal{F}_{tt}
    \end{pmatrix},
    \end{equation}
    where the off-diagonal elements \(\mathcal{F}_{gt}=\mathcal{F}_{tg}\) capture the correlations between the estimations of \(g\) and \(t\). This matrix provides a comprehensive measure of the information content available from the measurement.
The corresponding Cram\'er-Rao bound for $g$, including the effect of the nuisance parameter $t$ is then given by
    \begin{equation}\label{eq:CR_general}
    \mathrm{Var}(g) \;\ge\; \bigl(\mathbf{\mathcal{F}}^{-1}\bigr)_{gg} \;=\; \frac{\mathcal{F}_{tt}}{\mathcal{F}_{gg}\,\mathcal{F}_{tt} \;-\; \bigl[\mathcal{F}_{gt}\bigr]^2}\,.
    \end{equation}
    This bound sets the ultimate precision limit for any unbiased estimator of\(g\), accounting for correlations with time uncertainty.
    
    Since \(p(x_\lambda;g,t)\) depends on \(g\) only through \(\mathcal{C}_{n,n'}\) and \(\mathcal{F}_{n,n'}\), the derivative with respect to \(g\) can be expressed as 
    \begin{equation}\label{eq:dpdg_general}
    \partial_g \bigl[\mathcal{C}_{n,n'}\,\mathcal{F}_{n,n'}\bigr] \;=\; (\partial_g \mathcal{C}_{n,n'})\,\mathcal{F}_{n,n'} \;+\; \mathcal{C}_{n,n'}\,(\partial_g \mathcal{F}_{n,n'}).
    \end{equation}
    This decomposition shows that the sensitivity of the probability density to the parameter \(g\) has two contributions: one from the direct phase dependence in \(\mathcal{C}_{n,n'}\) and one from the mechanical overlap \(\mathcal{F}_{n,n'}\).
    
    Examining the derivative of \(\mathcal{C}_{n,n'}\) with respect to \(g\), we obtain
    \begin{equation}
    \partial_g \mathcal{C}_{n,n'} \;=\; \frac{\partial}{\partial g}\exp\!\Bigl\{-\,2\,i\,\bar{k}\,\bar{g}\,(n-n')\,\zeta(t)\Bigr\}
    \;=\; -\,2\,i\,A\,\bar{k}\,(n-n')\,\zeta(t)\, \mathcal{C}_{n,n'},
    \end{equation}
    since \(\partial_g(\bar{g})=A\). This derivative quantifies how a change in the coupling parameter \(g\) affects the phase evolution of the optical field.
        Similarly, writing \(\mathcal{F}_{n,n'}=\exp\{\mathcal{B}_{n,n'}\}\) leads to
    \begin{equation}
    \partial_g \mathcal{F}_{n,n'} \;=\; \mathcal{F}_{n,n'}\,\partial_g \mathcal{F}_{n,n'}.
    \end{equation}
    The derivative of the exponent \(\mathcal{B}_{n,n'}\) with respect to \(g\) is given by
    \begin{equation}\label{eq:dFdg_final}
{\partial_g \ln \mathcal{F}_{n,n'}(g,t)=0}
\qquad\text{(equivalently, }\partial_g \mathcal{F}_{n,n'}(g,t)=0\text{)}.
\end{equation}

    This result reflects the impact of the coupling on the mechanical overlap, and hence on the interference between different quantum pathways.
    
    Combining these derivatives, we define the auxiliary quantity {
    \begin{equation}\label{eq:Q_nn_def_g}
  Q^g_{n,n'} \;\equiv\; -2\,i\,A\,\bar{k}\,(n-n')\,\zeta(t).
 \end{equation}}
    This quantity \(Q^g_{n,n'}\) encapsulates the full sensitivity of the combined optical and mechanical contributions to variations in \(g\). 
    The derivative of the probability density with respect to \(g\) becomes
    \begin{equation}
    \partial_g p(x_\lambda;g,t) \;=\; e^{-|\alpha|^2}\,\sum_{n,n'} \mathcal{C}_{n,n'}\,\mathcal{F}_{n,n'}\,d_{n,n'}(x_\lambda)\,Q^g_{n,n'},
    \end{equation}
    and dividing by \(p(x_\lambda;g,t)\) yields the logarithmic derivative \(\partial_g\ln p\), a key ingredient in computing the Fisher information.
    
    A completely analogous procedure applies for differentiation with respect to the time parameter \(t\). Since \(t\) appears both in the phase factor of \(\mathcal{C}_{n,n'}\) (through \(\zeta(t)\)) and in the overlap exponent \(\mathcal{F}_{n,n'}\), we define
    \begin{equation}
    Q^t_{n,n'}(t,g)
    \;\equiv\; \frac{\partial_t\bigl[\mathcal{C}_{n,n'}(t,g)\,\mathcal{F}_{n,n'}(t,g)\bigr]}{\mathcal{C}_{n,n'}(t,g)\,\mathcal{F}_{n,n'}(t,g)}.
    \end{equation}
    After differentiating the contributions from both \(\mathcal{C}_{n,n'}\) and \(\mathcal{F}_{n,n'}\), the final expression is found to be
  {\begin{equation}\label{eq:Qt_final}
\begin{aligned}
Q^t_{n,n'}(t,g)
=\;&
i\Bigl[\bar{k}^2\bigl(n^2-n'^2\bigr)-2\bar{k}\bar{g}\,(n-n')\Bigr]\,(1-\cos t)
\\[4pt]
&\;+\;
\Bigl[-\bar{k}^2(n-n')^2+2i\,\bar{k}(n-n')\,\beta_I\Bigr]\sin t ,
\end{aligned}
\end{equation}}

The function \(Q^t_{n,n'}(t,g)\) quantifies the sensitivity of the measurement outcome to temporal variations, incorporating both the effects of phase evolution and the modifications in the mechanical overlap due to time evolution.
    
    Thus, the time derivative of the probability density is given by
    \begin{equation}
    \partial_t p(x_\lambda;g,t) \;=\; e^{-|\alpha|^2}\,\sum_{n,n'} \mathcal{C}_{n,n'}\,\mathcal{F}_{n,n'}\,d_{n,n'}(x_\lambda)\,Q^t_{n,n'},
    \end{equation}
    and dividing by \(p(x_\lambda;g,t)\) yields the logarithmic derivative \(\partial_t\ln p\).
    
    Using the definition
    \begin{equation}
    \partial_\theta \ln p(x_\lambda;g,t) \;=\; \frac{\partial_\theta p(x_\lambda;g,t)}{p(x_\lambda;g,t)},
    \end{equation}
    the elements of the CFI matrix can be constructed as
    \begin{align}
    \mathcal{F}_{gg} &\;=\; \int dx_\lambda\, \frac{\bigl[\partial_g p(x_\lambda;g,t)\bigr]^2}{p(x_\lambda;g,t)}, \label{eq:Fgg_final}\\[1mm]
    \mathcal{F}_{tt} &\;=\; \int dx_\lambda\, \frac{\bigl[\partial_t p(x_\lambda;g,t)\bigr]^2}{p(x_\lambda;g,t)}, \label{eq:Ftt_final}\\[1mm]
    \mathcal{F}_{gt} &\;=\; \int dx_\lambda\, \frac{\partial_g p(x_\lambda;g,t)\,\partial_t p(x_\lambda;g,t)}{p(x_\lambda;g,t)}. \label{eq:Fgt_final}
    \end{align}
  
    \begin{figure}
     \centering
        \includegraphics[width=1\linewidth]{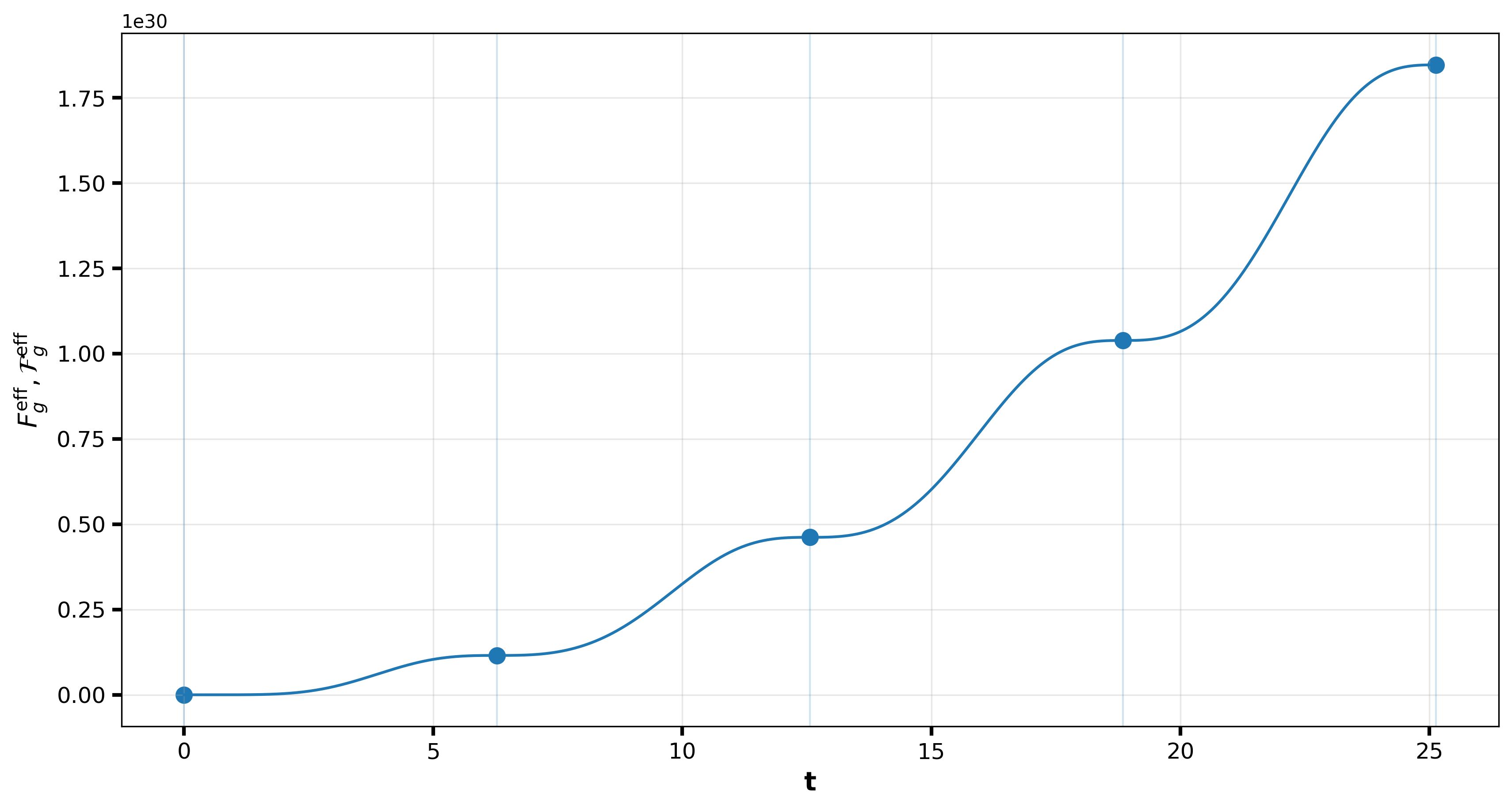}
         \caption{{Comparison} between CFI at stroboscopic times  and QFI  for the optomechanical parameters listed in Fig. \ref{fig1}.}
   \label{fig2}
    \end{figure}
  
    These elements quantify the information content in the measurement regarding the parameters \(g\) and \(t\). In particular, \(\mathcal{F}_{gg}\) and \(\mathcal{F}_{tt}\) measure the sensitivity with respect to each parameter individually, while the off-diagonal element \(\mathcal{F}_{gt}\) accounts for the correlations between the two.
    
    Alternatively, the CFI matrix elements can be expressed in a “sum-over-\(n,n'\)” form by substituting the expressions for \(\partial_g p\) and \(\partial_t p\). In this form, the elements become
    \begin{equation}
    \mathcal{F}_{gg} \;=\; e^{-|\alpha|^2}\,\int dx_\lambda\;
    \frac{\bigl[\sum_{n,n'} \mathcal{C}_{n,n'}\,\mathcal{F}_{n,n'}\,d_{n,n'} \,Q^g_{n,n'}\bigr]^2}{\sum_{n,n'} \mathcal{C}_{n,n'}\,\mathcal{F}_{n,n'}\,d_{n,n'}},
    \end{equation}
    \begin{equation}
    \mathcal{F}_{tt} \;=\; e^{-|\alpha|^2}\,\int dx_\lambda\;
    \frac{\bigl[\sum_{n,n'} \mathcal{C}_{n,n'}\,\mathcal{F}_{n,n'}\,d_{n,n'} \,Q^t_{n,n'}\bigr]^2}{\sum_{n,n'} \mathcal{C}_{n,n'}\,\mathcal{F}_{n,n'}\,d_{n,n'}},
    \end{equation}
    \begin{equation}
    \mathcal{F}_{tg} \;=\; e^{-|\alpha|^2}\,\int dx_\lambda\;
    \frac{\bigl[\sum_{n,n'} \mathcal{C}_{n,n'}\,\mathcal{F}_{n,n'}\,d_{n,n'} \,Q^t_{n,n'}\bigr]
        \bigl[\sum_{n,n'} \mathcal{C}_{n,n'}\,\mathcal{F}_{n,n'}\,d_{n,n'} \,Q^g_{n,n'}\bigr]}{\sum_{n,n'} \mathcal{C}_{n,n'}\,\mathcal{F}_{n,n'}\,d_{n,n'}}.
    \end{equation}
	
These expressions involve long sums over Fock indices and an integral over the continuous quadrature variable $x_\lambda$. Because of this complexity, assessing the optimality of homodyne detection is not immediate and requires numerical evaluation. 
In Fig.~\ref{fig2} we directly compare the QFI and CFI at stroboscopic times, where optimality has been demonstrated in the single-parameter case~\cite{NonlinearGravimetry2018}. We observe that homodyne detection remains optimal at such stroboscopic times, even in the presence of time uncertainty. By contrast, at other evolution times the QFI and CFI differ even in the single-parameter case~\cite{NonlinearGravimetry2018}. The SLD formalism guarantees the existence of a measurement and estimator that saturate the QFI, both in single- and multi-parameter settings. However, for generic times the corresponding optimal measurement may be nontrivial in practice and could require adaptive strategies, such as dynamically adjusting the local oscillator phase in response to prior outcomes. Exploring such adaptive schemes in the presence of time uncertainty remains an open direction for future work.

	\section{Conclusions} \label{section_V} 
    
In this work, we have derived the ultimate sensitivity limits for a broad class of quantum sensors, when time is treated as an intrinsically uncertain parameter. We have shown that the coupling between time and parameter estimation imposes a more stringent constraint on the achievable estimation precision, with the standard {Heisenberg} limit reachable only under specific, ideal conditions. We also discussed how this limit connects to the fundamental bounds on quantum evolution, clarifying the origin of precision loss as a direct manifestation of the Mandelstam-Tamm timescale. These general results were then applied to a prototype optomechanical gravimeter, for which we explicitly constructed the full Fisher information matrix and calculated the QCRB.
To date, experimental implementations of optomechanical gravimeters have demonstrated noise-equivalent acceleration sensitivities of $\sim 1.4\mu $g$/\sqrt{Hz}$ in a photonic crystal optomechanical cavity \cite{Krause2012}, and a sensitivity of $\sim$ 0.1 $\mu$g \cite{Yang2025} both limited by thermal noise of the mechanical oscillator. However these systems, particularly those based on levitated nanoparticles, could potentially reach quantum-limited performances, achieving acceleration sensitivities of $\sim 10^{-16} $g \cite{NonlinearGravimetry2018,qvarfort2021optimal}, thereby surpassing the best atomic interferometers even for low optical intensities. Experiments achieving these performances would not only enable measurements of gravitational acceleration at unprecedented levels, but could also directly probe the effects of intrinsic time uncertainty on parameter estimation.

	

	\bibliographystyle{apsrev4-1}
	\bibliography{references}

@article{RevModPhys.92.015004,
  title = {Sensitivity optimization for NV-diamond magnetometry},
  author = {Barry, John F. and Schloss, Jennifer M. and Bauch, Erik and Turner, Matthew J. and Hart, Connor A. and Pham, Linh M. and Walsworth, Ronald L.},
  journal = {Rev. Mod. Phys.},
  volume = {92},
  issue = {1},
  pages = {015004},
  numpages = {68},
  year = {2020},
  month = {Mar},
  publisher = {American Physical Society},
  doi = {10.1103/RevModPhys.92.015004},
  url = {https://link.aps.org/doi/10.1103/RevModPhys.92.015004}
}

@article{bonus,
  author    = {Bonus, F. and Knapp, C. and Valahu, C.H and Mironiuc, M. and Weidt, S. and Hensinger, W. K.},
  title     = {Ultrasensitive single-ion electrometry in a magnetic field gradient},
  journal   = {Nature Physics},
  volume    = {21},
  pages     = {1189–1195},
  year      = {2025}
}

@article{gilmore,
author = {Kevin A. Gilmore  and Matthew Affolter  and Robert J. Lewis-Swan  and Diego Barberena  and Elena Jordan  and Ana Maria Rey  and John J. Bollinger },
title = {Quantum-enhanced sensing of displacements and electric fields with two-dimensional trapped-ion crystals},
journal = {Science},
volume = {373},
number = {6555},
pages = {673-678},
year = {2021},
doi = {10.1126/science.abi5226}
}

@article{JunoGravity,
  author    = {Iess, L. and Folkner, W. M. and Durante, D. and Parisi, M. and Kaspi, Y. and Galanti, E.},
  title     = {Measurement of Jupiter’s asymmetric gravity field},
  journal   = {Nature},
  volume    = {555},
  pages     = {220--222},
  year      = {2018},
  doi       = {10.1038/nature25776}
}

@article{tinoreview,
  author    = {Tino, G. M.},
  title     = {Testing gravity with cold atom interferometry: results and prospects},
  journal   = {Quantum Sci. Technol.},
  volume    = {6},
  pages     = {024014},
  year      = {2021},
  doi       = {10.1088/2058-9565/abd83e}
}

@article{kasevich,
  author    = {Asenbaum, P. and Overstreet, C and. Kim, M. and Curti, J and Kasevich, M. A.},
  title     = {Atom-interferometric test of the equivalence principle at the $10^{−12}$ level},
  journal   = {Phys. Rev. Lett.},
  volume    = {125},
  pages     = {191101},
  year      = {2020},
  doi       = {10.1103/PhysRevLett.125.191101}
}

@article{Rosi2014,
  author    = {Rosi, G. and Sorrentino, F. and Cacciapuoti, L. and Prevedelli, M. and Tino, G. M.},
  title     = {Precision measurement of the Newtonian gravitational constant using cold atoms},
  journal   = {Nature},
  volume    = {510},
  pages     = {518--521},
  year      = {2014},
  doi       = {10.1038/nature13433}
}

@article{Montenegro2025, 
  title = {Heisenberg-limited spin-mechanical gravimetry},
  author = {Montenegro, Victor},
  journal = {Phys. Rev. Res.},
  volume = {7},
  issue = {1},
  pages = {013016},
  numpages = {15},
  year = {2025},
  month = {Jan},
  publisher = {American Physical Society},
  doi = {10.1103/PhysRevResearch.7.013016},
  url = {https://link.aps.org/doi/10.1103/PhysRevResearch.7.013016}
}

@article{Magnetomechanics,
  author    = {Geraci, A. A. and Weinstein, D.},
  title     = {Proposed magnetically levitated test mass for gravity measurements},
  journal   = {Physical Review D},
  volume    = {80},
  number    = {10},
  pages     = {104016},
  year      = {2009},
  doi       = {10.1103/PhysRevD.80.104016}
}

@article{Armata,
  title = {Quantum limits to gravity estimation with optomechanics},
  author = {Armata, F. and Latmiral, L. and Plato, A. D. K. and Kim, M. S.},
  journal = {Phys. Rev. A},
  volume = {96},
  issue = {4},
  pages = {043824},
  numpages = {6},
  year = {2017},
  month = {Oct},
  publisher = {American Physical Society},
  doi = {10.1103/PhysRevA.96.043824},
  url = {https://link.aps.org/doi/10.1103/PhysRevA.96.043824}
}

@article{GRACE1,
  author    = {R. D. Ray and S. B. Luthcke},
  title     = {Tide model errors and GRACE gravimetry: towards a more realistic assessment},
  journal   = {Geophysical Journal International},
  volume    = {167},
  number    = {3},
  pages     = {1055},
  year      = {2006}, 
  doi       = {10.1111/j.1365-246X.2006.03229.x}
}

@article{GRACE2,
  author    = {J. L. Chen and C. R. Wilson and B. D. Tapley},
  title     = {Satellite Gravity Measurements Confirm Accelerated Melting of Greenland Ice Sheet},
  journal   = {Science},
  volume    = {313},
  number    = {5795},
  pages     = {1958},
  year      = {2006}, 
  doi       = {https://doi.org/10.1126/science.1129007}
}

@article{AtomInterferometry1,
  author    = {A. Peters and K. Y. Chung and S. Chu},
  title     = {High-precision gravity measurements using atom interferometry},
  journal   = {Metrologia},
  volume    = {38},
  number    = {1},
  pages     = {25},
  year      = {2001},
  doi       = {10.1088/0026-1394/38/1/4}
}

@article{Magnetomechanics1,
  author    = {M. Johnsson and G. Brennen and J. Twamley},
  title     = {Macroscopic superpositions and gravimetry with quantum magnetomechanics},
  journal   = {Scientific Reports},
  volume    = {6},
  pages     = {37495},
  year      = {2016},
  doi       = {10.1038/srep37495}
}

@article{NonlinearGravimetry2018,
  author    = {S. Qvarfort and A. Serafini and P. F. Barker and others},
  title     = {Gravimetry through non-linear optomechanics},
  journal   = {Nature Communications},
  volume    = {9},
  pages     = {3690},
  year      = {2018},
  doi       = {10.1038/s41467-018-06037-z}
}

@book{Kay1993,
  author = {Kay, Steven M.},
  title = {Fundamentals of Statistical Signal Processing: Estimation Theory},
  publisher = {Prentice Hall},
  year = {1993},
  address = {Englewood Cliffs, NJ},
  isbn = {978-0133457117}
}

@book{LehmannCasella1998,
  author = {Lehmann, Erich L. and Casella, George},
  title = {Theory of Point Estimation},
  publisher = {Springer-Verlag},
  year = {1998},
  edition = {2nd},
  isbn = {978-0387985022},
  doi = {10.1007/b98854}
}

@article{Braunstein1994,
  author = {Braunstein, Samuel L. and Caves, Carlton M.},
  title = {Statistical distance and the geometry of quantum states},
  journal = {Physical Review Letters},
  volume = {72},
  number = {22},
  pages = {3439--3443},
  year = {1994},
  publisher = {American Physical Society},
  doi = {10.1103/PhysRevLett.72.3439}
}

@article{Giovannetti2011,
  author = {Giovannetti, Vittorio and Lloyd, Seth and Maccone, Lorenzo},
  title = {Advances in quantum metrology},
  journal = {Nature Photonics},
  volume = {5},
  number = {4},
  pages = {222--229},
  year = {2011},
  publisher = {Nature Publishing Group},
  doi = {10.1038/nphoton.2011.35}
}

@article{Paris2009,
  author = {Paris, Matteo G. A.},
  title = {Quantum Estimation for Quantum Technology},
  journal = {International Journal of Quantum Information},
  volume = {7},
  number = {supp01},
  pages = {125--137},
  year = {2009},
  publisher = {World Scientific},
  doi = {10.1142/S0219749909004839}
}

@article{ciufolini2016,
  author = {Ciufolini, I. and et al},
  title = {A test of general relativity using the lares and lageos satellites and a grace earth gravity model},
  journal = {The European Physical Journal C},
  volume = {76},
  pages = {120},
  year = {2016},
  doi = {10.1140/epjc/s10052-016-3961-8}
}

@article{fuentes14,
  author = {Ahmadi, M. and Bruschi, D. E. and Fuentes, I.},
  title = {Quantum metrology for relativistic quantum
fields},
  journal = {Phys. Rev. D},
  volume = {98},
  pages = {065028},
  year = {2014},
  doi = {10.1103/PhysRevD.89.065028}
}

@article{fuentes19,
  author = {Bruschi, D. E. and Datta, A. and Ursin, R. and Ralph, T. C. and Fuentes, I.},
  title = {Quantum-metrology estimation of spacetime
parameters of the earth outperforming classical precision},
  journal = {Phys. Rev. A},
  volume = {99},
  number = {065028},
  year = {2019},
  doi = {10.1103/PhysRevA.99.032350}
}

@article{fuentes23,
  author = {Bravo, T. and R\"atzel, D. and Fuentes, I.},
  title = {Gravitational time dilation in extended quantum sys-
tems: The case of light clocks in Schwarzschild space-
time},
  journal = {AVS Quantum Science},
  volume = {5},
  pages = {014401},
  year = {2023},
  doi = {10.1116/5.0123228}
}

@article{mazumdar12,
  author = {Biswas, T. and Gerwick, E. and Koivisto, T. and Mazumdar, A.},
  title = {Towards singularity-and ghost-
free theories of gravity},
  journal = {Phys. Rev. Lett.},
  volume = {108},
  pages = {031101},
  year = {2012},
  doi = {10.1103/PhysRevLett.108.031101}
}

@article{BECInterferometry,
  author = {Hardman, K. S. and Everitt, P. J. and McDonald, G. D. and Manju, P. and Wigley, P. B. and Sooriyabadara, M. A. and Kuhn, C. C. N. and Debs, J. E. and Close, J. D. and Robins, N. P.},
  title = {A quantum sensor: simultaneous precision gravimetry and magnetic gradiometry with a Bose-Einstein condensate},
  journal = {Physical Review Letters},
  volume = {117},
  number = {13},
  pages = {138501},
  year = {2016},
  doi = {10.1103/PhysRevLett.117.138501}
}

@article{Lewandowski2021,
  author = {Lewandowski, Charles W. and Knowles, Tyler D. and Etienne, Zachariah B. and D'Urso, Brian},
  title = {High-Sensitivity Accelerometry with a Feedback-Cooled Magnetically Levitated Microsphere},
  journal = {Physical Review Applied},
  volume = {15},
  number = {1},
  pages = {014050},
  year = {2021},
  doi = {10.1103/PhysRevApplied.15.014050}
}

@article{Monteiro2017,
  author = {Monteiro, Fernando and Ghosh, Sumita and Fine, Adam Getzels and Moore, David C.},
  title = {Optical levitation of 10-ng spheres with nano-g acceleration sensitivity},
  journal = {Physical Review A},
  volume = {96},
  number = {6},
  pages = {063841},
  year = {2017},
  doi = {10.1103/PhysRevA.96.063841}
}

@article{qvarfort2021optimal,
  title={Optimal estimation of time-dependent gravitational fields with quantum optomechanical systems},
  author={Qvarfort, Sofia and Serafini, Alessio and Xuereb, Andr{\'e} and Braun, Daniel and R{\"a}tzel, Dennis and Bruschi, David Edward},
  journal={Physical Review Research},
  volume={3},
  number={1},
  pages={013159},
  year={2021},
  publisher={APS}
}

@article{Krause2012,
  title        = {A high‐resolution microchip optomechanical accelerometer},
  author       = {Krause, Alexander G. and Winger, Martin and Blasius, Tim D. and Lin, Qiang and Painter, Oskar},
  journal      = {Nature Photonics},
  volume       = {6},
  pages        = {768--772},
  year         = {2012},
  doi          = {10.1038/nphoton.2012.245},
}

@article{Yang2025,
  title        = {From Photon Momentum Transfer to Accelerometer Based on Optical Levitated Microsphere at Dynamic Input from 0.1 μg to 1 g},
  author       = {Yang, Jianyu and Li, Nan and Pan, Yuyao and Yang, Jing and Chen, Zhiming and Cai, Han and Wang, Yuliang and Han, Chuankun and Chen, Xingfan and Liu, Cheng and Hu, Hui-Zhu},
  journal      = {ACS Photonics},
  volume       = {12},
  pages        = {194--201},
  year         = {2025},
  doi          = {10.1021/acsphotonics.4c01477},
}

@article{Asenbaum2020, 
  title = {Atom-Interferometric Test of the Equivalence Principle at the ${10}^{\ensuremath{-}12}$ Level},
  author = {Asenbaum, Peter and Overstreet, Chris and Kim, Minjeong and Curti, Joseph and Kasevich, Mark A.},
  journal = {Phys. Rev. Lett.},
  volume = {125},
  issue = {19},
  pages = {191101},
  numpages = {5},
  year = {2020},
  month = {Nov},
  publisher = {American Physical Society},
  doi = {10.1103/PhysRevLett.125.191101},
  url = {https://link.aps.org/doi/10.1103/PhysRevLett.125.191101}
}

@book{Helstrom1976,
  author    = {C. W. Helstrom},
  title     = {Quantum Detection and Estimation Theory},
  publisher = {Academic Press},
  year      = {1976},
}

@article{MandelstamTamm1945,
  author  = {L.~Mandelstam and I.~Tamm},
  title   = {The Uncertainty Relation Between Energy and Time in Non‐Relativistic Quantum Mechanics},
  journal = {J.~Phys.\ (USSR)},
  volume  = {9},
  pages   = {249--254},
  year    = {1945},
}

@article{lind,
  author  = {Lindkvist, Joel and Sab\'in, Carlos and Johanson, G. and Fuentes, Ivette},
  title   = {Motion and gravity effects in the precision of quantum clocks},
  journal = {Scientific Reports},
  year    = {2019},
  volume  = {5},
  number  = {4},
  pages   = {10070}
}

@article{ahmadi,
  author  = {Ahmadi, M. and Bruschi, D. E. and Sab\'in, Carlos and Adesso, Gerardo and Fuentes, Ivette},
  title   = {Relativistic Quantum Metrology: Exploiting relativity to improve quantum measurement technologies},
  journal = {Scientific Reports},
  year    = {2014},
  volume  = {4},
  pages  = {4996}
}

@article{colombo,
  author  = {Colombo, Simone and Pedrozo-Penarel, Edwin and Vuleti\'c, Vladan},
  title   = {Entanglement-enhanced optical atomic clocks},
  journal = {Applied Physics Letters},
  year    = {2022},
  volume  = {121},
  pages   = {210502}
}

@article{zwierz,
  author  = {Zwierz, M. and P\'erez-Delgado, C. A. and Kok, P.'},
  title   = {General Optimality of the Heisenberg Limit for Quantum Metrology},
  journal = {Physical Review Letters},
  year    = {2010},
  volume  = {105},
  pages   = {180402}
}

@article{Suzuki2020Nuisance,
  author       = {Jun Suzuki},
  title        = {Quantum estimation theory with nuisance parameters},
  journal      = {Journal of Physics A: Mathematical and Theoretical},
  year         = {2020},
  volume       = {53},
  number       = {26},
  pages        = {264001},
  doi          = {10.1088/1751-8121/ab8b78},
  eprint       = {2001.03942},
  archivePrefix= {arXiv},
  primaryClass = {quant-ph}
}

@article{SidhuKok2020Multiparameter,
  author       = {Jasminder S. Sidhu and Pieter Kok},
  title        = {Geometric perspective on quantum parameter estimation},
  journal      = {AVS Quantum Science},
  year         = {2020},
  volume       = {2},
  number       = {1},
  pages        = {014701},
  doi          = {10.1116/1.5119961},
  eprint       = {1912.08405},
  archivePrefix= {arXiv},
  primaryClass = {quant-ph}
}
	
\end{document}